# Optoelectronic sampling of ultrafast electric transients with single quantum dots

Alex Widhalm[1,2,3,4], Sebastian Krehs[1,3,4], Dustin Siebert[2,3], Nand Lal Sharma[1,a], Timo Langer[1,3,4], Björn Jonas[1,3,4], Dirk Reuter[1,3,4], Andreas Thiede[2,3], Jens Förstner[2,3,*], and Artur Zrenner[1,3,4,†]

[1] Paderborn University, Physics Department, Warburger Straße 100, 33098 Paderborn, Germany
[2] Paderborn University, Electrical Engineering Department, Warburger Straße 100, 33098 Paderborn, Germany
[3] Paderborn University, Center for Optoelectronics and Photonics Paderborn (CeOPP), Warburger Straße 100, 33098 Paderborn, Germany
[4] Paderborn University, Institute for Photonic Quantum Systems, Warburger Straße 100, 33098 Paderborn, Germany
[a] present address: Institute for Integrative Nanosciences, Leibniz IFW Dresden, Helmholtzstraße 20, 01069 Dresden, Germany

*jens.foerstner@upb.de
†artur.zrenner@upb.de

*In our work, we have engineered low capacitance single quantum dot photodiodes as sensor devices for the optoelectronic sampling of ultrafast electric signals. By the Stark effect, a time-dependent electric signal is converted into a time-dependent shift of the transition energy. This shift is measured accurately by resonant ps laser spectroscopy with photocurrent detection. In our experiments, we sample the laser synchronous output pulse of an ultrafast CMOS circuit with high resolution. With our quantum dot sensor device, we were able to sample transients below 20 ps with a voltage resolution in the mV-range.*

The field of quantum sensing has attracted tremendous interest during the last decades. The use of quantum systems as sensors promises high sensitivity, high precision, and access to nanoscale applications [1]. Today, quantum sensors detect a wide range of physical quantities such as magnetic fields, electric fields, temperature, and even gravity. With our present work we contribute to the field of electric field sensing. Here, existing concepts for quantum sensing rely mainly on diamond NV-centers with a special focus on the nanoscale [2], and on Rydberg atoms for RF-field sensing [3].

In this contribution, we use single semiconductor quantum dots (QDs) and ps laser techniques to demonstrate QD-based electric field sensing with ultrafast time-resolution. At low temperatures, QD exciton ground state transitions have extremely narrow line width [4] and excellent coherence properties, allowing for coherent state control [5, 6]. Integrated in diode structures, their transition energies are electric field tunable with high precision via the Stark effect. For resonant excitation, single QD photodiodes act as voltage tunable spectrometers, which also allow for electric read-out by photocurrent detection [7, 8]. We have engineered radio frequency (RF) compatible single QD Schottky-photodiodes as sensor devices for the sampling of ultrafast electric signals applied to the Schottky-gate. The optoelectronic sampling concept relies on resonant ps laser excitation and tracing of the QD exciton resonance via a back-gate voltage for each time step. For the demonstration of optoelectronic sampling we used the laser synchronous output pulse of an ultrafast CMOS circuit, which was connected to the Schottky-gate of the QD photodiode.

For the experimental realization of QD-based optoelectronic sampling we have fabricated photodiodes from MBE grown material containing InGaAs QDs with a low areal density in the range of 0.1 µm$^{-2}$, as required for single QD applications. The QDs are embedded in an intrinsic GaAs layer with 360 nm width, between a buried n$^+$ layer and a semitransparent titanium Schottky gate. The Schottky-diode was designed as low capacitance device to achieve high speed operation (see Ref. [9] for details). The QD sensor structure fabricated for this work has an active Schottky gate area of 120 µm$^2$, resulting in a diode capacitance of about 50 fF.

As signal source for the QD sensor we used ultrafast cryogenic CMOS electronics. For this purpose, we designed a customized chip based on a 0.13 µm SiGe:C BiCMOS technology from the IHP Leibniz institute [10]. The electric output pulse $V_P(t)$ generated by the CMOS chip is variable in amplitude and polarity by means of the supply voltage $V_{DD}$ and the control voltage $V_{Sign}$ (see Fig 1 (c)). The functionality of the chip is given by a set of logical gates for trigger pulse recovery and pulse polarity switching, which finally drive an output stage with 27 paralleled inverters to keep the output impedance low. The system integration was realized on a ceramic chip carrier. To allow for high-speed operation, the GaAs QD photodiode chip and the CMOS chip were mounted in close proximity and connected electrically by short distance wire bonding (see Fig1 (a)). For efficient broadband decoupling of the supply and control pads of the chip, we used RF single layer capacitor chips. The arranged system was operated at T=4.2 K.

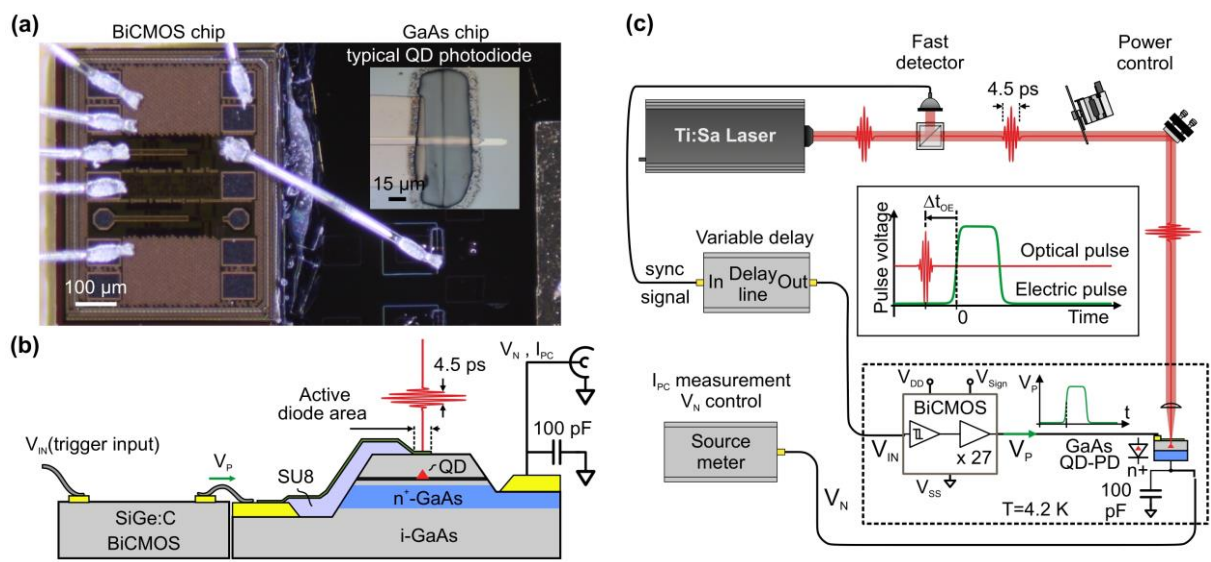

Fig. 1. (a) Photograph of the integrated BiCMOS chip and low capacitance QD photodiode. A detailed view of the photodiode is shown in the inset. (b) Cross sectional view of the CMOS driver chip and the low-capacitance QD photodiode. (c) Experimental setup for QD based optoelectronic sampling. BiCMOS chip and low capacitance photodiode are operated at T=4.2 K. A fast photodetector is used as trigger source for the BiCMOS chip. A variable delay-line controls the timing between the optical and electric pulses.

Fig. 1 (c) illustrates the experimental setup required for the implementation of QD based optoelectronic sampling. A mode-locked Ti:sapphire laser generates Fourier transform limited 4.5 ps optical pulses with a repetition rate of 80 MHz. Part of the laser output is directed to a fast photo receiver with 12 GHz bandwidth to generate an electric synchronization signal. By means of an adjustable power attenuator, the remaining laser output is adjusted to an intensity, which drives the QD exciton transition with a pulse area of 0.75·π. This pulse area was selected to compromise between signal to noise ratio and the oscillatory Rabi regime above π. Finally, the laser beam is directed to a cryogenic microscope setup (T= 4.2 K) and focused on the QD photodiode by an objective lens with NA=0.75.

The electric output of the fast photo receiver drives a tunable delay-line (see Fig. 1 (c)), which precisely controls the time delay between the electric and the ps optical pulses over a range of 5 ns with a resolution of less than 1 ps. We refer to this delay as optoelectronic delay $\Delta t_{OE}$ in the following text. The width of the electric pulse is extended to a duration of about 1.3 ns by pulse stretching. As a result of this, the timing jitter of the falling edge of the pulse is significantly higher (about 20 ps) than the leading edge (about 6 ps). Finally, the electric pulse is applied to the input of the CMOS chip in the cryostat via a semi-rigid coaxial line. The output voltage $V_P(t)$ of the CMOS chip is connected to the anode (Schottky-gate) of the low capacitance photodiode by short distance wire bonding (see Fig 1 (a) and (b)). A source meter is connected to the cathode ($n^+$ contact) of the photodiode, which is RF-grounded by a 100 pF capacitor. The source meter supplies the cathode voltage $V_N$ to the photodiode and detects its photocurrent $I_{PC}$ with high sensitivity and low noise (30 fA/$\sqrt{Hz}$). The central energy of the ps laser is tuned to resonance with the QD exciton ground state transition when $V_P(t)=V_{SS}=0$ V (i.e. $V_P$ in the logical "low" condition). The maximum of the photocurrent resonance peak is then observed at $V_N = 1.1$ V. Further, we have taken care to realize equal propagation times of the electric and optical pulses on their way to the QD photodiode. Therefore, the incoming electric pulse is synchronized with exactly the same optical pulse that leads also to the buildup of population in the QD.

The optoelectronic sampling method employed here is based on the Stark effect acting on the energy of the QD exciton ground state transition. By the Stark effect, a time-dependent electric field is converted into a time-dependent shift of the transition energy. With high exciton energies in the near infrared range this correspondence holds even in the ultrafast regime as demonstrated with THz electric fields [11]. In the past, electric field dependent resonance fluorescence was already used for the time resolved detection of transitions and charging effects in single QDs [12].

In the current work, we expose a single QD in a low capacitance diode structure to a time dependent electric field, generated by an ultrafast CMOS chip. The chip supplies the pulsed external bias voltage $V_P(t)$ to the anode (Schottky gate) of the photodiode (see Fig. 2).

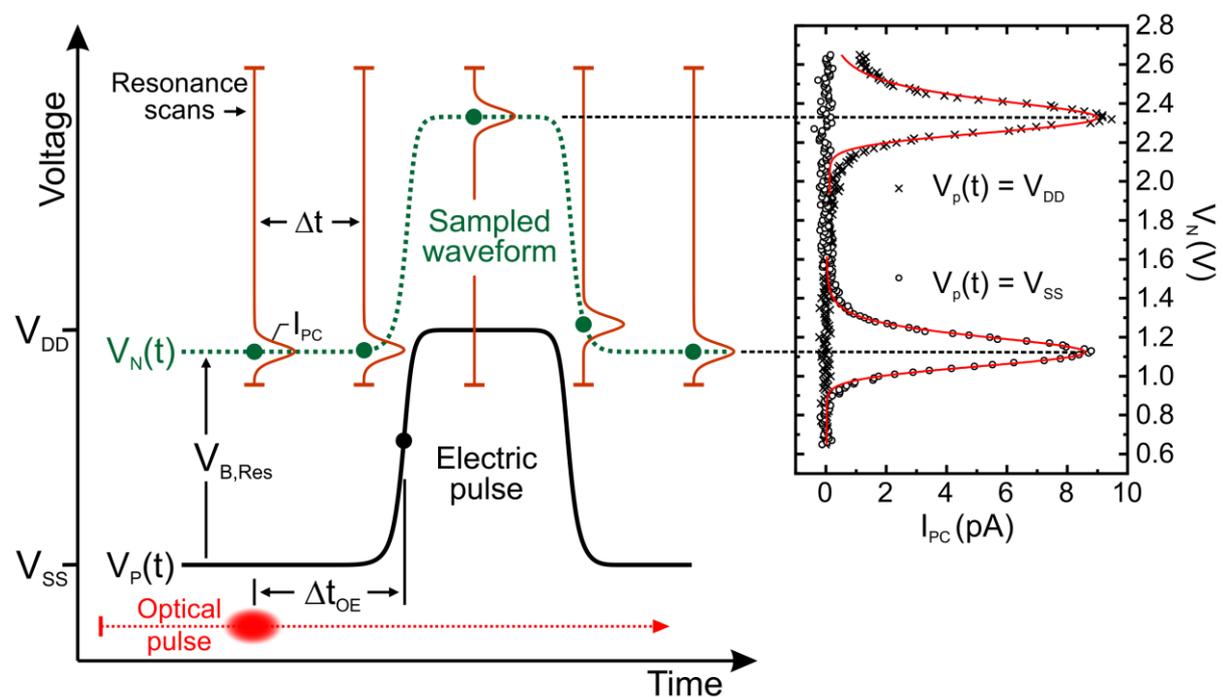

Fig. 2. Illustration of QD-based optoelectronic sampling. The electric pulse to be sampled has a voltage swing from $V_{SS}=0$ V to $V_{DD}=1.2$ V. Performing resonance scans for different optoelectronic delays $\Delta t_{OE}$, the sampled waveform can be determined from the center positions of the photocurrent resonance peaks. Experimental data for resonance scans at two different pulse voltages are shown on the right.

To sample $V_P(t)$ we measure the time-dependent shift of the QD transition energy by resonant ps photocurrent spectroscopy. Basically, such scans can be performed along the voltage or the time axis. In the current work we keep the laser energy fixed and perform resonance scans by sweeping the cathode voltage $V_N$ for a set of optoelectronic delays $\Delta t_{OE}$ (separated by $\Delta t$), as indicated in Fig. 2. For the selected and fixed laser energy, the resonance occurs at the constant reverse bias condition $V_{B,Res}=V_P-V_N$. Therefore, the sampled waveform $V_N(t)$ appears vertically shifted from $V_P(t)$ by the constant displacement $V_{B,Res}$. The laser energy is selected such, that the reverse bias $V_{B,Res}$ is strong enough to allow for efficient photocurrent extraction. The resonance scans provide a photocurrent peak for each selected $\Delta t_{OE}$.

Basic theoretical considerations and numerical simulations show, that the position of the maximum of the photocurrent gives a very exact measure of $V_P$. Other aspects that one might assume to have large influence such as the width of the optical pulse, voltage-dependent tunneling rates, nonlinear scaling of the photocurrent, and rapid adiabatic passage effects only lead to a broadening and asymmetry of the photocurrent curve, but have only little effect on the accuracy as only the maximum position is relevant. Timing jitter is a negligible factor as long as it is below a certain threshold ($\approx$ quarter rise time), above that it can even lead to somewhat steeper sampled edges. This is illustrated in Fig. 3 showing very good agreement between an electric input pulse with 15 ps RC rise time, and a sampled pulse resulting from a numerical simulation of the sampling procedure assuming a jitter in the optoelectronic delay of 6.5 ps. Details are given in the supplementary materials.

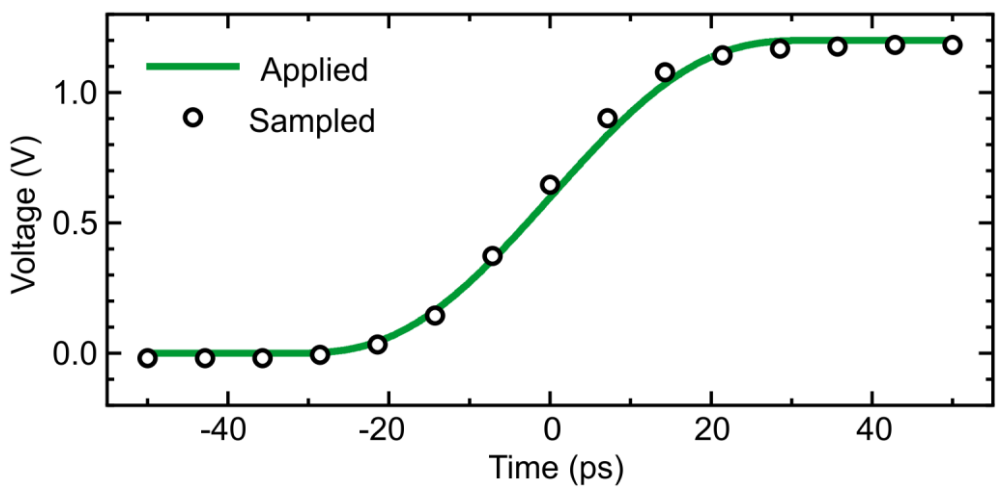

Fig. 3. Numerical calculations showing the excellent agreement between an applied electric pulse transient with an RC rise time of 15 ps (green solid curve), and the voltages obtained using the presented sampling method (black circles) assuming a timing jitter of 6.5 ps. Neither the optical pulse width, voltage-dependent tunneling rates, nonlinearities, nor timing jitter (if below a certain threshold) significantly affect the accuracy.

To reliably obtain the maximum position in the experiment, exponentially modified Gaussian distribution (exGaussian) fits are applied to the scanned asymmetric resonances giving accurate values for the resonant cathode voltage $V_N$. In the inset of Fig. 2 we show experimental data and fits for two different sampling points, which apply for $V_P(t)=V_{SS}$ and $V_P(t)=V_{DD}$. For a complete acquisition of the sampled waveform, the described procedure must be performed for a sufficiently dense series of optoelectronic delays $\Delta t_{OE}$. In Fig. 4 we show the sampled waveform of an input pulse with 1.3 ns width and 1.2 V amplitude, delivered by our CMOS chip. For the acquisition of the waveform we incremented the optoelectronic delay $\Delta t_{OE}$ in steps of 7 ps. The time zero point of $\Delta t_{OE}$ was set to the leading edge of the pulse. The obtained data shows a highly resolved sampling result, both in time and amplitude. The leading edge of the pulse has a finite rise-time, which is mainly a result of the p-MOS channel resistance in the CMOS chip and the QD diode capacitance. In the left inset of Fig. 4 we show a detailed view of the leading edge. Over a wide voltage range, the sampled transient follows nicely an exponential RC fit function with a time constant of 16.7 ps (shown in red).

Even a small capacitive feed-through effect from the gates to the drains of the output inverters can be resolved at t≈-30 ps, in quantitative agreement with a simulation performed with the Cadence Virtuoso design kit (see Fig. 4, inset on the right). Here, the gate drive of the MOS transistors results in a negative displacement current on the drain side and hence in a transient voltage undershoot of $V_P(t)$ [13]. After the rising edge, the sampled voltage contains ringing contributions from the entire circuitry, which the fast leading edge of the pulse excites. All ringing effects are damped out before the next sampling event occurs (repetition time 12.5 ns). The falling edge of the pulse is subjected to stronger timing jitter as compared to the leading edge. We discuss the details of this later.

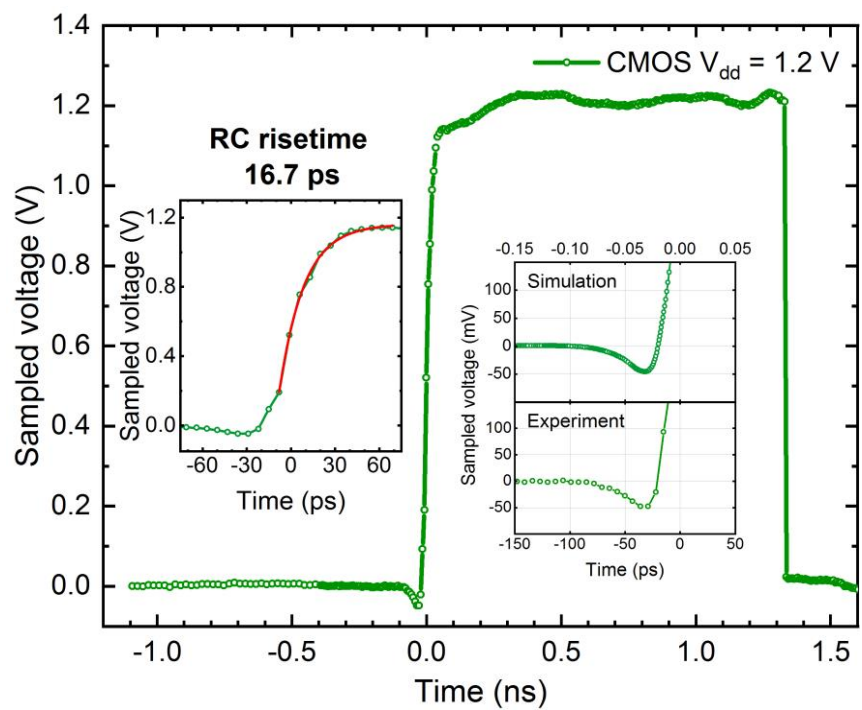

Fig. 4. Electric pulse form obtained by QD based optoelectronic sampling. Inset left: Expanded view of the rising pulse edge and comparison to a RC model. Inset right: Detailed view of the experimentally observed transient voltage undershoot caused by a capacitive feed-through effect compared to a Cadence simulation.

The voltage resolution of our sampling method depends on several parameters. It is affected by the integration time used in the photocurrent measurements and by the number of voltage steps used for the resonance scans. This determines the precision of the asymmetric Gaussian fit to the measured resonance curve and the accuracy of the obtained resonant cathode voltage $V_N$. For an integration time of 100 ms and a resonance scan voltage step width of 15 mV we obtained standard deviations of the exGaussian fit functions in the range of 1 mV. In regions with fast electric transients, the QD transition is electrically chirped [14]. Here, the resonance shape and amplitude change for different time positions along the electric transient (see Fig. 5 (a)). For $\Delta t_{OE}$ = −43 ps the sampled voltage remains almost constant. Here, the exGaussian fit function is almost symmetric and has a full width at half maximum (FWHM) of about $\Delta V_{exp}$ = 0.163 ± 0.002 V. This value is determined by the spectral width of the Fourier transform limited laser pulse with a temporal width of 4.5 ± 0.4 ps. Taking the time bandwidth product of 0.44 into account, this corresponds to a spectral FWHM of $\Delta E_{FWHM}$ = 0.41 ± 0.04 meV [15]. With the known Stark shift of the QD exciton transition, this results in a FWHM of the photocurrent resonance of $\Delta V_{4.5\ ps}$ = 0.156 ± 0.014 mV.

This value agrees very well with the measured value of $\Delta V_{exp}$ = 0.163 ± 0.002 mV. For $\Delta t_{OE}$ = -1 ps, in the steepest section of the electric transient, the QD exciton transition is electrically chirped with 0.039 meV/ps. Here, we observe a much broader resonance peak. The standard deviation of the central resonance position obtained from exGaussian fits increases almost by a factor of three, from 0.82 mV without chirp to 2.2 mV with chirp. In Fig. 5 (a) we show experimental data for five resonance scans, covering the most interesting range across the rising edge region. The full set of sampled voltages versus time (step size 7 ps) is shown in Fig. 5 (b).

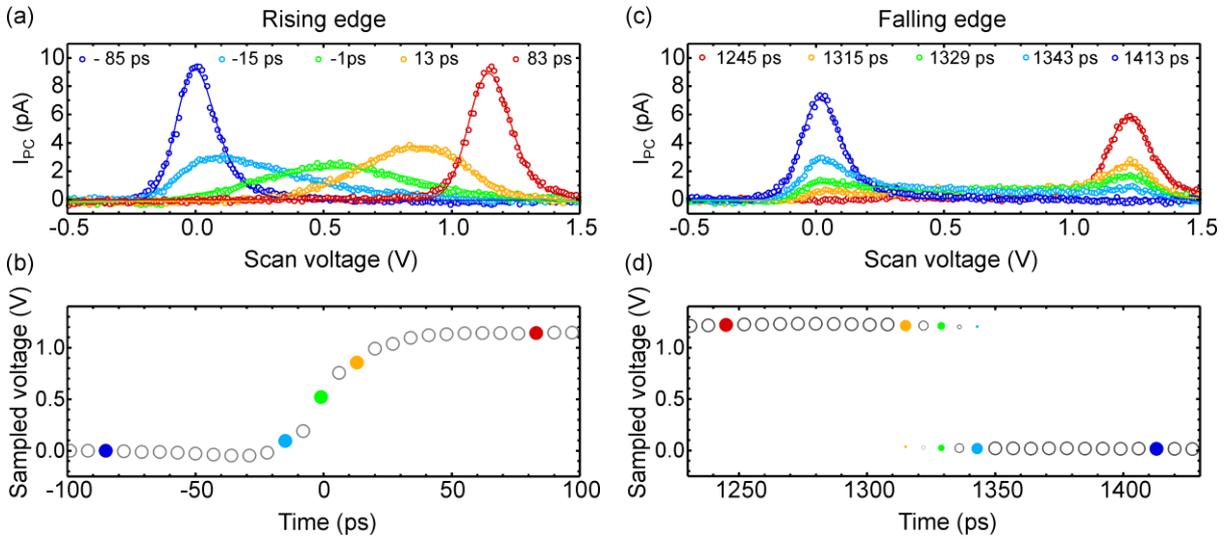

Fig. 5. (a) Photocurrent resonance scans with exGaussian fits obtained at the rising edge of the electric pulse. The line broadening in the transition region is caused by electric chirp acting on the QD transition. (b) Reconstructed waveform at the rising edge. (c) Resonance scans with exGaussian fits obtained at the falling edge. Due to enhanced timing jitter and fast transition, the resonances occur on either the high voltage or low voltage side. (d) Sampling process applied to the falling edge, two circles appear at some time values corresponding to two photocurrent maxima (circle radius reduced for small amplitudes).

Our sampling results for the rising edge of the electric pulse have been obtained for a scenario, where the observed electric RC pulse rise time is in the range of 20 ps, but the laser pulse width and the timing jitter are both in the regime between 5 ps and 7 ps. At the falling edge of the electric pulse the situation is different in two respects. First, the channel resistance of the n-MOS transistor is lower as compared to the p-MOS transistor and we expect a shorter transit time. Second, the timing jitter at the falling edge of the electric pulse is considerably larger as compared to the rising edge. The rising edge of the fast photo receiver (12 GHz bandwidth) controls directly the timing jitter of the rising edge of the electric pulse. The timing jitter between the ps laser pulse and the rising edge of the electric pulse is therefore very low (about 6 ps). The timing of the falling edge of the pulse is not precisely correlated to the leading edge, because the total length of the electric pulse has been stretched to about 1.3 ns. This leads to an increase of the timing jitter at the falling edge of the electric pulse to about 20 ps. In a repetitive sampling experiment this has the consequence, that in a transition region either the high level or the low level is sampled, but less likely the transition region in between. As a result, we obtain the resonance scenario shown in Fig. 5 (c), where either the high level or the low level resonance is observed with complementary probability when the jittered transition region is scanned with respect to the arrival time of the ps laser pulse. Nevertheless, the observed timing jitter remains limited to about 3 to 4 time steps of 7 ps. This becomes clear from Fig. 5 (d), were the probability for a high-level or low-level event is mapped in the area of the colored dots, as obtained from the area under the resonance curves shown in Fig. 5 (c). As shown in the supplement, both scenarios can be well explained by theoretical analysis based on the full quantum optoelectronic model introduced in Ref. 14.

In summary, we presented optoelectronic sampling of ultrafast electric signals with low capacitance single QD photodiodes as sensor devices. We use the Stark effect to convert a time-dependent electric signal into a time-dependent shift of the QD transition energy. For the time resolved measurement of this shift we perform resonant ps laser spectroscopy with spectrally tunable photocurrent detection. We were able to sample the laser synchronous output pulse of an ultrafast CMOS circuit exhibiting transients below 20 ps with a voltage resolution in the mV-range. The accuracy is not affected or limited by a moderate timing jitter, nor the spectral optical pulse width as theoretical calculations show. With the currently demonstrated performance, the optoelectronic sampling method appears extremely powerful, with high potential for further improvements and applications.

## AUTHORS' CONTRIBUTIONS

A.W. and S.K. contributed equally to the experimental part of this work.

## ACKNOWLEDGMENTS

The authors like to acknowledge financial support by the Deutsche Forschungsgemeinschaft (DFG, German Research Foundation), Projektnummer 231447078-TRR 142 (via project A03 and C04), the BMBF via the Q.Link.X project No. 16KIS0863; and the PC$^2$ for the computation time.

SUPPLEMENTARY MATERIAL

To analyze and validate the experimental findings, we applied the theoretical model developed in Ref. [14], which is based on the optical Bloch equations for a two-level system (TLS) extended by off-resonant dark states $\rho_e$ and $\rho_{hh}$ for a QD occupied by only one electron or hole [16]. Also, the voltage dependence of tunneling and dephasing as well as jitter is considered. This provides an accurate model to simulate the entire experimental setup. In Fig. S1 (a) the theoretical data on the rising edge is shown assuming a symmetric Gaussian distributed timing jitter in the optoelectronic delay $\Delta t_{OE}$ with a standard deviation of $\sigma = 6.5$ ps. Such a moderate jitter in combination with chirp can even lead to a slight sharpening of the sampled edge compared to the applied one, as visible in Fig. 3, since the jitter moves sampled voltages slightly further to the close-by upper/lower level. For even smaller jitter this effect vanishes completely.

We see that our theoretical results explain the experimental characteristics very well. For an optoelectronic delay scale similar to the experimental one, the peak for $\Delta t_{OE} = 0$ ps in the middle of the rising edge is reduced and broadened due to timing jitter influences and a chirped excitation of the QD two level system [14]. This behavior vanishes the more the optoelectronic delay is set to the electric pulse's low and high voltage level regions.

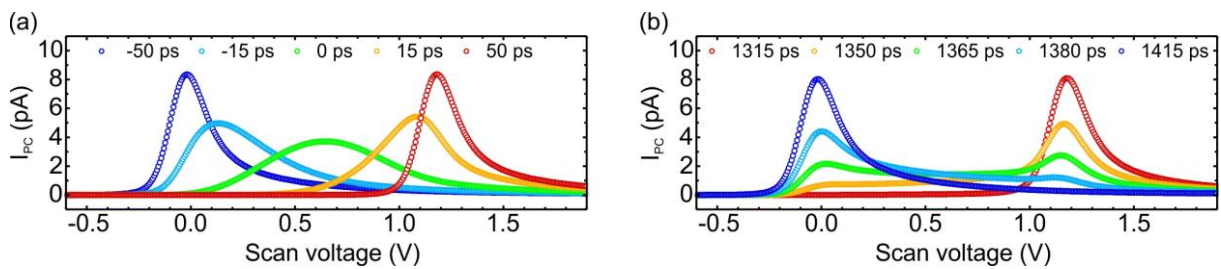

Fig. S1: (a) Theoretical optoelectronic sampling at the rising edge of the electric transient with an assumed timing jitter of $\sigma = 6.5$ ps. The optoelectronic delay $\Delta t_{OE}$ varies from the low voltage level (-50 ps) to the high voltage level (50 ps) of the electric pulse. (b) Sampling at the falling edge of the electric pulse with $\sigma = 20$ ps. Therefore, the optoelectronic delay $\Delta t_{OE}$ varies from the high voltage level (1315 ps) to the low voltage level (1415 ps).

For the falling edge we obtain the same characteristic result as in the experiment. Here, with $\sigma = 20$ ps and $\tau_{fall} = 12$ ps, the peak for $\Delta t_{OE} = 1365$ ps at the center of the falling edge is not present at the expected voltage $V_N + V_{B,Res} = 0.6$ V. Instead, we see two peaks appearing at the resonance voltages for the low and high level of the electric pulse caused by a steeper falling edge and a higher timing variance, which illustrates the sensitivity of this sampling method to the influence on jitter (see Fig. S1 (b)). Additionally, we can provide a full resampling of the applied electric pulse as shown in Fig. 3 of the main text for the rising edge. This helps us quantifying the sampling method of picking the maximum photocurrent for the resonance voltage. We demonstrate that the rule of taking the maximum photocurrent is appropriate for a precise reconstruction of the electric pulse.

Overall, our theoretical model can describe both characteristics of the rising and falling edge respectively. It confirms the experimental data and helps understanding impacts on the optoelectronic sampling of ultrafast electric transients.